\documentclass[aps,superscriptaddress,showpacs,amsmath,amssymb,amsfonts,altaffillsymbol,twocolumn,notitlepage,prl]{revtex4-2}

\usepackage{wasysym}
\usepackage{graphicx}
\usepackage{dcolumn}
\usepackage{bm}
\usepackage{color}
\usepackage[normalem]{ulem}

\begin{document}

\preprint{APS/123-QED}

\title{Criticality and marginal stability of the shear jamming transition of frictionless soft spheres}

\author{Varghese Babu}
\email{varghese@jncasr.ac.in}

\author{Srikanth Sastry} 
\email{sastry@jncasr.ac.in}
\affiliation{%
Jawaharlal Nehru Center for Advanced Scientific Research, Jakkur Campus, Bengaluru 560064, India.
}%
\date{\today}

\begin{abstract}
We study numerically the critical behavior and marginal stability of the shear jamming transition for frictionless soft spheres, observed to occur over a finite range of densities, associated with isotropic jamming for densities above the minimum jamming (J-point) density. Several quantities are shown to scale near the shear jamming point in the same way as the isotropic jamming point. We compute the exponents associated with the small force distribution and the interparticle gap distribution, and show that the corresponding exponents are consistent with the marginal stability condition observed for isotropic jamming, and with predictions of the mean field theory of jamming in hard spheres.
\end{abstract}

\maketitle



Jamming is a ubiquitous phenomenon, observed in a variety of physical systems classified as  {\it granular matter} (sand, grains, powders), foams,  colloids {\it etc}, wherein thermal motion does not play a significant role. The jamming transition is typically observed when disordered materials are compressed, resulting in a transition  to a state which can resist deformation \cite{liu1998jamming}. Athermal frictionless soft-spheres have served as an idealized model for studying the jamming transition \cite{OHern_2003,Hecke_2009,liu2010jamming}, along with packings of hard spheres. 

The jamming density $\phi_J$ has many properties of a critical point. The configurations at $\phi_J$ are isostatic with average co-ordination number $z = 2d$, where $d$ is the spatial dimension \cite{OHern_2003,Hecke_2009,liu2010jamming}. As the 
system is compressed above the jamming density (for soft-spheres) the excess co-ordination number scales 
as $\delta z \sim (\phi-\phi_J)^{\frac{1}{2}}$, independent of the interaction potential and the configurations are mechanically stable \cite{Hecke_2009,OHern_2003,liu2010jamming}.
The pressure vanishes linearly with $(\phi-\phi_J)$, with a pre-factor depending on the interaction potential \cite{OHern_2003,chaudhuri2010jamming,liu2010jamming}. Close to, and above, the jamming density, the vibrational density of states (VDOS) $D(\omega)$, where $\omega$ is the frequency, displays anomalous 
behavior, with the presence of excess low-frequency modes compared to the Debye solid which describe a normal elastic medium, with a characteristic frequency $\omega^{*}$ vanishing as the jamming point is approached, with a power law dependence on $(\phi-\phi_J)$ \cite{wyart2005rigidity,wyart2005effects,wyart2005geometric}.
Such aspects of critical-like behaviour near the jamming point has been widely investigated and established \cite{liu2010jamming,GoodrichPNAS2016}.  

The requirement of mechanical stability of jammed packings has been shown to imply an inequality, or bound, between exponents that characterise the distribution of inter-particle forces $f$, which exhibits a power law form $P_e(f) \sim f^\theta_e$ at small forces, and the distribution of distances, or gaps $h$, between particles that are nearly in contact, which exhibits a well known power-law singularity, $g(h) \sim h^{-\gamma}$ \cite{wyart2012marginal,lerner2013low}. The bound, $\gamma \geq 1/(2 + \theta_e)$, was argued to be saturated at jamming \cite{wyart2012marginal} and that such a {\it marginal stability condition} provided a mechanism to explain the avalanches of rearrangement observed \cite{combe2000strain,jin2017exploring}. 

The mean field theory of glass transition in hard spheres in the limit of infinite dimensions \cite{charbonneau2017glass,parisi2020theory}, interestingly, leads to predictions concerning the behavior at jamming, and in particular a prediction for the exponents $\theta$ and $\gamma$ to be $\theta = 0.42311..$ and $\gamma = 0.41269..$. While the predicted value of $\gamma$ is close to those observed in two (2D) and three dimensional (3D) packings, as well as higher dimensions \cite{charbonneau2012universal}, leading to the possibility that $d = 2$ constitutes the upper critical dimensions for the jamming transition, the reported values of exponent $\theta$ {\cite{degiuli2014force,lerner2013low,charbonneau2012universal}} exhibits a wide range. However, as noted in \cite{lerner2013low}, the presence of localised excitations leads to a modified distribution $P_l(f) \sim f^{\theta_l}$ and marginal stability condition $\gamma \geq (1 - \theta_l)/2$. The localised excitations were associated with sphere arrangements prone to buckling, or  {\it bucklers}, in \cite{charbonneau2015jamming}, and separating out the distributions of forces corresponding to bucklers leads to the verification of marginal stability condition, with the exponents predicted by mean field theory.

In this letter, we examine the validity of these aspects of criticality and marginal stability for the shear jamming transition for frictionless soft spheres in two and three dimensions, and show that they are indeed valid. The jamming of granular matter under shear  has been observed experimentally \cite{bi2011jamming,ren2013reynolds} and numerically \cite{vinutha2016disentangling,vinutha2019force,vinutha2020timescale} for frictional systems. However friction is not necessary for the shear jamming transition \cite{kumar2016memory,jin2018stability,baity2017emergent,babu2021dilatancy,jin2021jamming,peshkov2021critical}. In \cite{baity2017emergent}, critical behavior near the shear jamming transition for frictionless soft spheres were considered in 3D. Although the density range over which shear jamming occurs was seen to vanish in the thermodynamic limit, the behavior of the pressure, contact number $z$ and the bulk modulus were shown to exhibit the same behavior as a function of shear stress (equivalently, shear strain above the jamming strain) as at the isotropic jamming point. The key difference is that both the bulk modulus $B$ and the shear modulus $G$ remain finite at the shear jamming point, unlike the isotropic case where only the bulk modulus is finite. Nevertheless, only a single eigenvalue of the elastic modulus tensor becomes finite, and the finite values of $B$ and $G$ can be understood in terms of a rotation of the eigenvectors, leading to the conclusion that shear jamming and isotropic have the same symmetry and critical behavior. Similar conclusions have been arrived at in other investigations {\cite{urbani2017shear,jin2021jamming}} 

However, shear jamming can occur over a finite range of densities even in the absence of friction \cite{babu2021dilatancy,kumar2016memory,jin2018stability,jin2021jamming}, a possibility related to the presence of a range of densities (J-line) over which isotropic jamming can take place, above the minimum jamming density $\phi_J$ {(of  $\approx 0.648$ for $d=3$ and $\approx 0.84$ for $d=2$)}. When the preparation history dependent jamming density of a packing $\phi_j$ is $> \phi_J$, decompression of the packing to densities $\phi$, $\phi_J < \phi < \phi_j$ leads to unjamming, but such unjammed packings can jam under the application of shear \cite{babu2021dilatancy,jin2021jamming} (The jamming density $\phi_j \approx \phi_J$ in \cite{baity2017emergent}, and hence the finite shear jamming range is not observed). 

It is of interest therefore  investigate the critical behavior of shear jamming over such density intervals, which we do, building on previous work \cite{baity2017emergent,peshkov2021critical,vinutha2020timescale,jin2021jamming}. With the distance from the shear jamming strain $\gamma - \gamma_j$ playing the role of the excess density above jamming, $\phi - \phi_j$ for isotropic jamming, we find that the scaling of pressure, excess contact number and shear stress and the behavior of the VDOS $D(\omega)$ is the same as at isotropic jamming. In addition, we explore in detail the marginal stability condition, employing the approach of \cite{charbonneau2015jamming} to distinguish localised excitations, or {\it bucklers}, and demonstrate that the behavior for shear jamminng is consistent with that for isotropic jamming and mean field predictions. Other than a preliminary investigation in \cite{vinutha2019force}, the applicability of the marginal stability condition for shear jamming has not been investigated. Our results thus clearly demonstrate that properties related to criticality and marginal stability for shear jamming are the same as for isotropic jamming. 

\noindent{\it Model and Methods:} The systems we study are bi-disperse soft sphere  mixtures {($50:50$)} in 2D and 3D with a harmonic repulsive inter-particle potential. The interaction potential is given by $v(|\vec{r}_{ij}|)=\epsilon(1-\frac{|\vec{r}_{ij}|}{\sigma_{ij}})^2$ for 
$|\vec{r}_{ij}|\leq\sigma_{ij}$ where $\vec{r}_{ij}$ is the distance between particles $i$ and $j$ and 
$\sigma_{ij}=\frac{\sigma_i+\sigma_j}{2}$ with $\sigma_k$ being the diameter of particle type $k$ 
{$(\sigma_2/\sigma_1 = 1.4)$}. The first step in our study involves generating configurations with 
jamming density $\phi_j > \phi_J$. For this we follow the protocol similar to the one used in 
\cite{chaudhuri2010jamming} which we describe for $d=3$ briefly; other procedures that could be employed 
are outlined in \cite{Das10203,babu2021dilatancy}. At the packing fraction $\phi=0.5935$ we generate 
configurations by initializing particle centers randomly and performing an energy minimization to generate 
a configurations with no overlaps (A configuration with no overlap is considered unjammed). This 
configuration is treated as a configuration of hard-spheres and equilibrated by hard-sphere Monte Carlo 
simulation using  HOOMD \cite{anderson2016scalable,anderson2020hoomd}. We then compress the system in steps 
of $\delta \phi=10^{-4}$, performing an energy minimization after each compression. When the jamming density 
$\phi_j$ is crossed, the energy after minimization $e/N$ will be greater than $10^{-24}$. When the energy 
crosses a 
 threshold (here $\frac{e}{N}>10^{-7}$) we stop the compression and start decompressing the system with smaller steps of  $\delta \phi=10^{-5}$. During the decompression when we are able to minimize the energy to $\frac{e}{N}<10^{-24}$ we stop the process and identify the
jamming density.  The jamming densities obtained through the procedure is distributed around $\phi\approx0.661$, which depends on the density of the initial equilibrated fluid \cite{chaudhuri2010jamming}. From the configurations at $\phi_j$ we generate unjammed configurations at $\phi_J<\phi<\phi_j$ by scaling the volume.


These configurations are sheared uniformly using Athermal Quasi Static (AQS) shear to observe shear jamming at a strain $\gamma_j$, employing LAMMPS \cite{plimpton1995fast}. 
AQS shear for a strain step $\delta \gamma$ is carried out by performing an affine transformation 
$x_i \rightarrow x_i + \delta \gamma \times y_i ; y_i \rightarrow y_i; z_i \rightarrow z_i$ 
of coordinates followed by  energy minimization. We generate configurations close to the jamming strain $\gamma_j$ and identify $\gamma_j$ as follows: We increment strain in steps of $\delta \gamma=10^{-3}$ until $\frac{e}{N}>e_{thresh}=10^{-7}$, at which point we redefine the strain step and threshold energy as
$\delta \gamma\rightarrow-1\times \delta \gamma/10$ and $e_{thresh}\rightarrow e_{thresh}/10$.
The system is strained in the reverse direction until $\frac{e}{N}<10^{-20}$ where updates to $e_{thresh}$ and $\delta \gamma$ are implemented again. This procedure is stopped when
$\delta \gamma<10^{-6}$ and $\frac{e}{N}<10^{-20}$ and the system is being reverse strained. 

Using this procedure we are able to obtain configurations close to the jamming strain, but to study the marginal stability of the shear jamming transition we need to generate configurations that are {\it{just}} shear-jammed. Quantitatively this means the configurations has a single self-stress state, or the contact network has one unique force-balance solution \cite{charbonneau2015jamming}. For a given jammed configuration with $N_c$ contacts and $N$ particles which are not rattlers, the number of self-stressed 
states is given by  $N_{ss}=N_c-(N-1)d$ with periodic boundary conditions \cite{charbonneau2015jamming}. We observe that for large system sizes  configurations obtained using the SJ procedure to obtain shear jammed configurations are not close enough to jamming and have multiple self-stressed states. 

To obtain configurations with single self-stressed state we adapt the procedure described for isotropic jamming in \cite{charbonneau2015jamming} for shear jamming. Starting with a configuration at $d\gamma \equiv  (\gamma - \gamma_j) = 10^{-5}$ as determined through the SJ procedure, we iteratively reduce the strain by exploiting the scaling of the potential energy $U\sim (\gamma-\gamma_j)^2$ (which is indeed observed, as shown in the Supplemental Material (SM) \footnote{See Supplemental Material at https://xxx for the additional information on:  
    i) Details of the iterative procedure (IP), 
    ii) Derivation of the expression for elastic constants,
    iii) Fabric anisotropy of the shear jammed configuration,
    iv) Results for pressure and contact number for 2D and 3D
    v) Size dependence of the shear jamming line {\it{vs}.} density and finite size scaling of the gap distribution, which includes 
    additional references \cite{lemaitre2006sum,goodrich2014jamming}}, in addition to further details of this procedure.) Using this procedure we generate shear-jammed configurations with a single 
self-stressed state, whose structure and forces we analyze to study the marginal stability condition. 
We follow similar procedures for the data regarding the isotropic case, with density instead of the strain as the control variable. 

The components of the stress tensor are is calculated using $\sigma_{\alpha \beta}  =\frac{1}{V} \sum_{i<j} f^{\alpha}_{ij} r^{\beta}_{ij}$, where $f_{ij}$ are the interparticle forces, and the pressure as $P = {1\over 3} \operatorname{tr}({\bf \sigma})$. We calculate the density of states $D(\omega)$ which is the distribution of $\omega=\sqrt{\lambda}$ where $\lambda$'s are the eigenvalues of the Hessian, for configurations over a range of strains above the jamming strain. As in the case of isotropic jamming we observe a plateau in $D(\omega)$ for small $\omega$, before $D(\omega)$ decreases to zero as $\omega \rightarrow 0$. The  frequency $\omega^*$ at which $D(\omega)$ decreases to half the plateau value is identified as the crossover frequency. 

We note that the jammed configurations analysed contain {\it rattlers}, particles with less than $d+1$ contacts. 
We remove rattlers recursively, by identifying them in each iteration from the configurations till no rattlers remain. The percentage of rattler particles is approximately $0.05\%$ for the cases considered. The average contact number, as well as distribution of gaps and forces reported, are obtained after the rattlers are removed. However, the packing fractions we report are calculated with the total number of particles. 


\begin{figure}
    \centering
    {\includegraphics[width=0.5\textwidth]{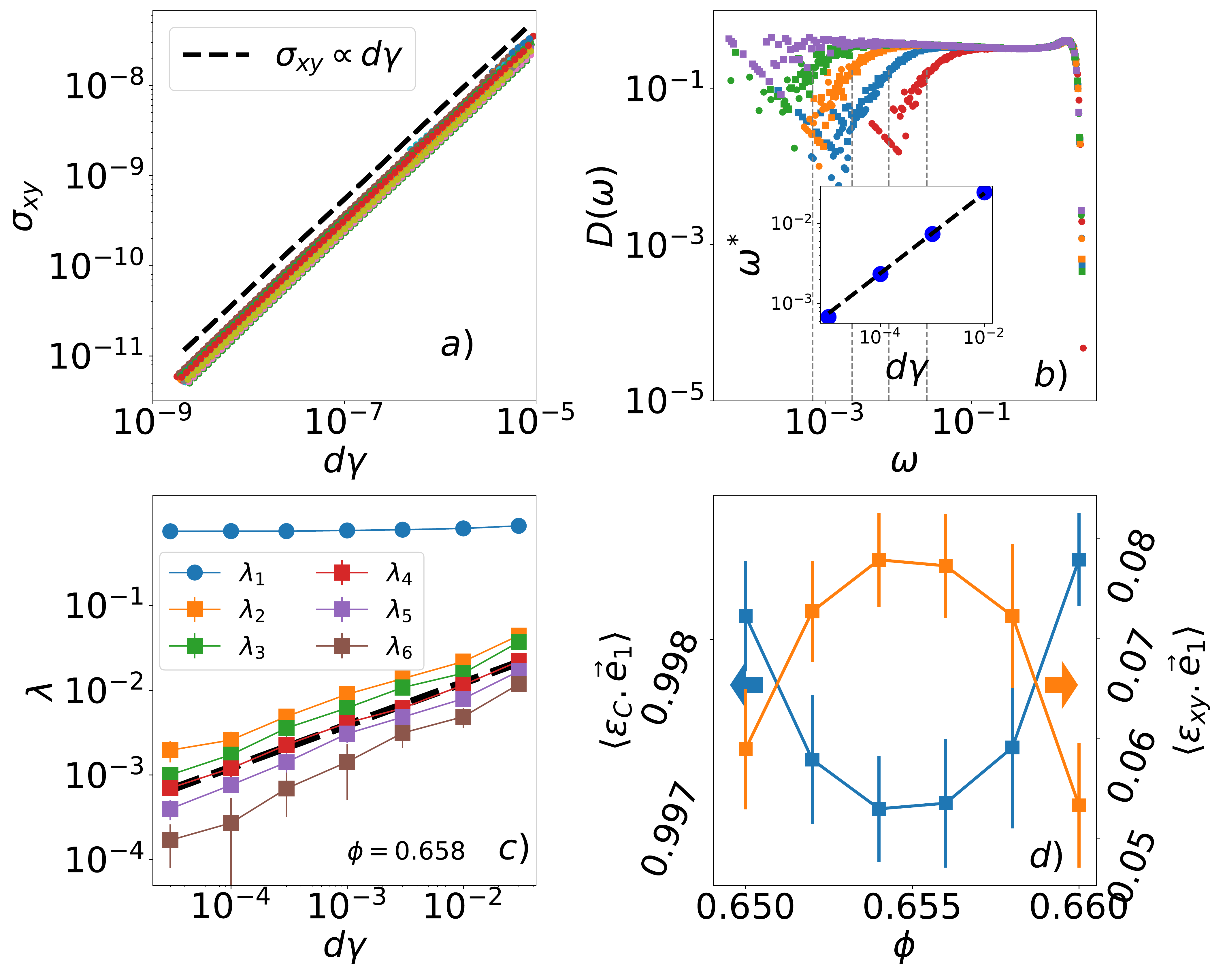}}

    \caption { 
    \textbf{a)} {Stress $\sigma_{xy}$ vs $d\gamma=\gamma-\gamma_j$ obtained by IP with fit line $\sigma_{xy} \sim d\gamma$.} \textbf{b)} VDOS $D(\omega)$ of configurations 
    at various $d\gamma$. The green, orange, blue and red symbols represent $d\gamma=10^{-5},10^{-4},10^{-3},10^{-2}$ respectively. The square and circle symbols represent  $\phi=0.658,0.656$ respectively. The violet curve is $D(\omega)$ calculated for isotropically jammed configurations with $N_{SS}=1$. \textbf{Inset:} The cross-over frequency $\omega^{*}$ is calculated 
    by choosing the frequency at which $D(\omega)$ becomes approximately half of the plateau value. These values are marked in \textbf{b)}. The spaced line shows the scaling $\omega^*\sim d\gamma ^{\frac{1}{2}}$. 
    \textbf{c)} Eigenvalues of the stiffness matrix $\lambda_i$ as a function of $d\gamma$ (for $N=8192$). One eigenvalue is significantly larger than the others and this corresponds to the bulk and shear modulus $C_{xyxy}$. The dashed line denotes an exponent of $1\over 2$. {$\lambda_i > \lambda_{i+1}$}
    \textbf{d)} Inner product of the eigenvector corresponding to the largest eigenvalue $\vec{e}_1$ of the stiffness matrix (for $d\gamma=10^{-5}$) with compressive (blue) and shear (orange) strain directions. }
    \label{fig1}
\end{figure}

\noindent{\it Results:} In Fig. \ref{fig1}(a), we show the variation of the shear stress $\sigma_{xy}$ vs $d\gamma$ for 3D, demonstrating linear behavior $\sigma_{xy} \sim d\gamma$ above the jamming strain. The pressure exhibits the same linear behavior and the excess contact number $\delta z = z-z_c$ varies with the distance from the jamming strain as $\delta z \sim \sqrt{d\gamma}$, with $z_c = 2d$, as observed for isotropic jamming. These results are shown in the SM, along with the corresponding results for 2D. In Fig.  \ref{fig1}(b), we show the VDOS $D(\omega)$, which exhibits a plateau at low frequencies corresponding to excess modes, which extend towards zero frequency as the jamming strain is approached from above. The frequency at which the crossover to the  
plateau occurs, $\omega^* \sim \delta \phi ^{\frac{1}{2}}$ for isotropic jamming \cite{wyart2005rigidity,wyart2005geometric}, and we observe the same scaling near the shear jamming transition, 
as shown in Fig. \ref{fig1}(b)(inset). In Fig. \ref{fig1}(c), we show the eigenvalues of the stiffness matrix (details of whose calculation are provided in the SM), investigated in \cite{baity2017emergent} for shear jamming in fricionless packings. As in \cite{baity2017emergent}, we find a nearly constant largest eigenvalue that is finite at the shear jamming point, and five nearly degenerate (but less so than in  \cite{baity2017emergent}) eigenvalues which are zero at shear jamming, and whose magnitude grows roughly as $d \gamma^{1/2}$ for larger strains. In \ref{fig1}d) we show the overlap of the eigenvector corresponding to the largest eigenvalue with the bulk strain direction and the shear strain direction, as a function of density $\phi$. Interestingly, the overlap of shear strain with the stiffest eigenvector shows a non-monotonic behavior. This is apparently related to the anisotropy of the configurations at shear jamming, quantified by the fabric anisotropy, which also shows a similar non-monotonic behavior with changing $\phi$, as shown in SM, and also observed in \cite{jin2021jamming}. These results taken together show that the nature of criticality near the shear jamming point is the same as that near the isotropic jamming point.

\begin{figure}[h]
    \centering
    {\includegraphics[width=0.48\textwidth]{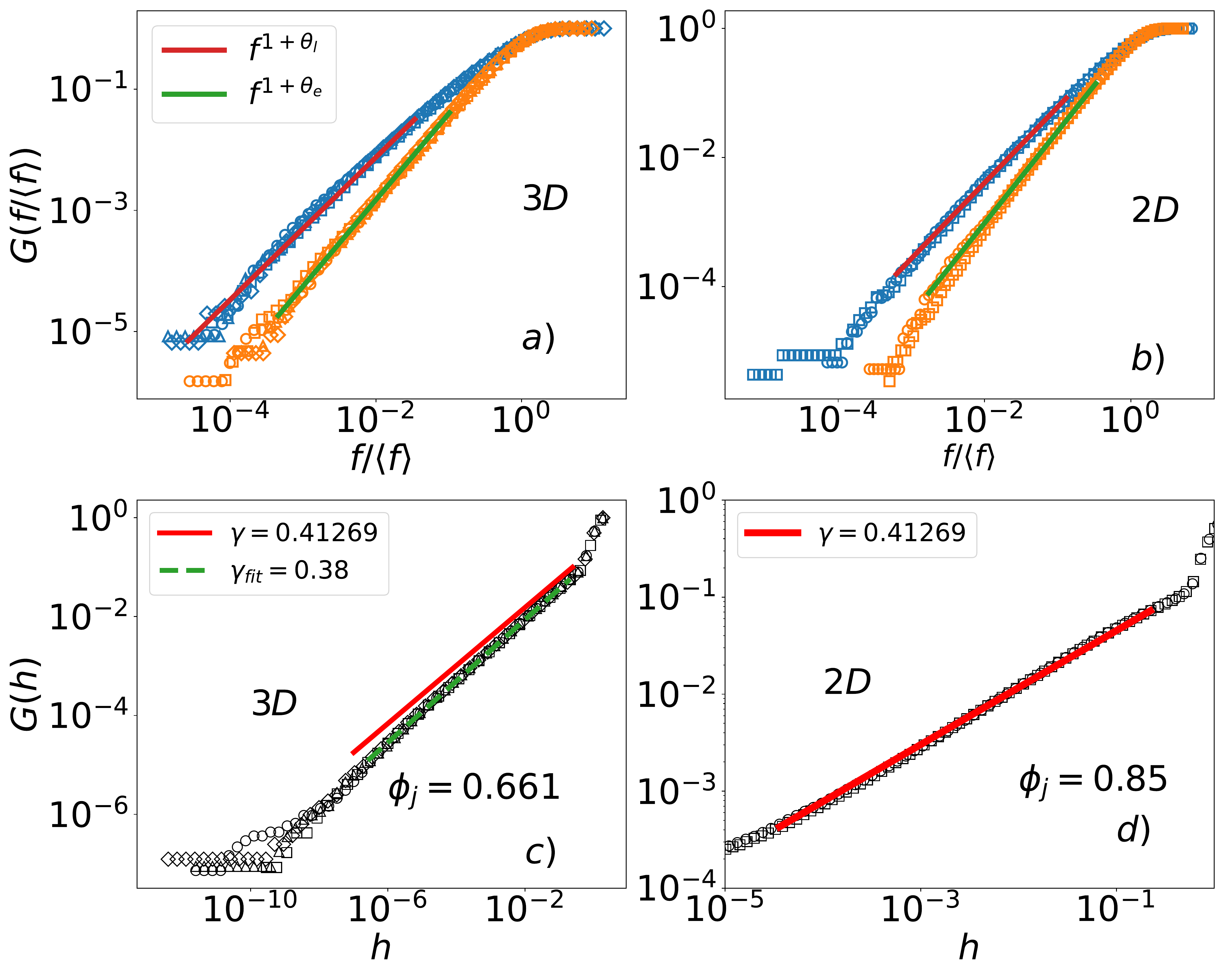}}
    \caption{Inter-particle forces and gaps in shear jammed configurations and comparison with iso-tropically jammed configurations. All configurations analyzed has a single self-stressed state.
    Symbols $\Diamond,\Circle,\Box,\triangle$ represent isotropic compression,$\phi=0.660,0.658,0.656$ respectively in $3D$ and symbols $\Circle,\Box$ represent $\phi=0.8499,0.8485$ in $2D$.
    \textbf{a)-b)} The cumulative distribution of forces $G(h)$. The blue symbols represent localized forces and the orange symbols represent the  extended forces. Comparison with exponents obtained from mean-field theory is shown. 
    \textbf{c)-d)} The cumulative distribution of gaps $h$. The red line shows the exponent from mean field theory.}
    \label{fig:forces_and_gaps}
\end{figure}

We now describe the results regarding the forces and the structure of the shear jammed 
configurations. It is convenient to consider the cumulative probability of forces, $G(f)=\int_{0}^{f}P(f')df'$. 
With the gap defined as $h=\frac{r-\sigma}{\sigma}$, the cumulative probability of gaps is $G(h)=\int_{0}^{h}P(h')dh'$. For isotropic jamming, the cumulative probability for forces (normalised to the mean value) is described by a power law $G(f/\langle f\rangle)\sim f^{1 + \theta}$ and for gaps, $G(h) \sim h^{1 - \gamma}$. As shown in \cite{lerner2013low} small forces in the system can
be either mechanically isolated, {\it i. e.} opening a contact will result only
in local rearrangement of the contact network, or localized modes,  or they can correspond to extended modes. 
Small forces which correspond to localised modes have a distribution characterized by exponent $\theta_{l}$ and
the forces corresponding to extended modes are characterized by
exponent $\theta_{e}$. The inequalities discussed in \cite{lerner2013low} in the two cases are, 

\begin{equation}
\gamma\ge\frac{1}{2+\theta_{e}} ~~; ~~ \gamma\geq\frac{1-\theta_{l}}{2}\label{eq:ineq}
\end{equation}

If the jammed system is marginally stable and the above inequalities Eq. \eqref{eq:ineq} are saturated (if they become equalities) then the following relation holds: 
\begin{equation}
\theta_{e}\gamma=\theta_{l}\label{eq:sat}
\end{equation}
The force-distribution calculated by including all the forces
in the system is characterized by exponent $\theta=\min(\theta_{l},\theta_{e}$). 
In order to extract the $\theta_l$ and $\theta_e$, we have to identify contacts associated with localized and extended modes correctly. Although the mean-field theory of hard-sphere glasses does not contain a prediction for $\theta_l$, based on the predicted values $\theta_e=0.42311..$ and 
$\gamma=0.41268..$ and Eq. \ref{eq:sat}, one has $\theta_l=0.17462$. Charbonneau {\it et al.}  \cite{charbonneau2015jamming} explored how to identify the contacts which carry small mechanically isolated forces. The mechanically isolated contacts are associated with ``buckler" particles which are particles with $d+1$ contacts. As shown in \cite{charbonneau2015jamming} the force distribution
calculated by including only the bucklers, $P_{l}(f)$, exhibits an exponent of $\theta_{l}=0.17462$. The force exponent calculated by using the distribution of the remaining forces is $P_{e}(f)$
shows an exponent of $\theta_{e}=0.42311$. We follow the same procedure to analyse configurations with a single self-stressed state identified by $N_c=(N-1)d+1$. As opposed to isotropic jamming, for the shear jamming transition we need take into account the effect of shear while classifying bucklers. However, we observe that for configurations at small-strains, classification of bucklers as particles with $d+1$ contacts is sufficient to obtain meaningful results. The cumulative probabilities of forces, separately for bucklers (localised modes) and the rest (extended modes), shown in Fig. \ref{fig:forces_and_gaps} (a) and (b) for 3D and 2D, show that indeed, the predicted values of $\theta_e$ and $\theta_l$ describe the data extremely well. 

In Fig. \ref{fig:forces_and_gaps} (c) and (d) we show the distribution of gaps for 3D and 2D. For 3D, while we find the mean field prediction of  $\gamma = 0.41268..$ closely describes the data, a value of $\gamma = 0.38$ is a better description of the data. Indeed, results in several works \cite{lerner2013low,charbonneau2012universal,jin2021jamming,charbonneau2021finite}, both for isotropic and shear jamming, are consistent with such a smaller exponent, which would correspond to a weak violation of the stability condition. However, the role of finite size effects in the observed departures at very small gaps has recently been investigated \cite{charbonneau2021finite}, emphasizing that finite size effects are much more pronounced for gaps rather than forces. A scaling collapse over several orders of magnitude are described supporting the accuracy of the mean field exponent for three dimensional packings. Our analysis of finite size effects for shear jamming, shown in the SM, clearly support the same conclusion. On the other hand, the results for 2D, shown in \ref{fig:forces_and_gaps} (d) agree very well with the mean field predictions. Thus, we conclude 
that the marginal stability conditions (Eq. \ref{eq:ineq}) is indeed valid for shear jamming, as well as they do for isotropic jamming, with the same universal exponents. 

In summary, we have numerically analysed configurations of soft spheres in two and three dimensions accurately generated at the shear jamming point, and above, for densities below the density $\phi_j$ at which they exhibit isotropic jamming, but above the minimum isotropic jamming density $\phi_J$. We show that several quantities, such as the pressure $P$, the excess contact number, and a crossover frequency $\omega^{*}$ in the VDOS $D(\omega)$ exhibit critical scaling that is identical to that at the isotropic jamming point, with the shear stress in addition displaying the same scaling as the pressure. We confirm the behavior of the eigenvalues of the stiffness matrix which have been investigated \cite{baity2017emergent} to argue that shear jamming has the same symmetry as isotropic jamming, and show that the rotation of the eigenvector of the largest eigenvalue in the shear strain direction is correlated with the anisotropy of the shear jammed structures. We show that the marginal stability condition is met for shear jamming to the same degree as for isotropic jamming, with exponents predicted by the mean field theory of the glass transition and jamming in hard spheres (although our results indicate that better finite size analysis is warranted for the gap distribution). Our results thus strongly support the idea that shear jamming displays the same critical behavior and marginal stability as isotropic jamming.

We acknowledge support from the Thematic Unit of Excellence on Computational Materials Science (TUE-CMS) and the National Supercomputing Mission facility (Param Yukti) at the Jawaharlal Nehru Centre for Advanced Scientific Research (JNCASR), for computational resources. SS acknowledges support through the JC Bose Fellowship (JBR/2020/000015) from the Science and Engineering Research Board, Department of Science and Technology, India.

 \bibliographystyle{apsrev4-2}
\bibliography{marginal_stability}

\end{document}